\documentclass[12pt,letterpaper]{article} 

\usepackage[a4paper, total={7in, 10in}]{geometry} 

\usepackage[super,comma,sort&compress]{natbib} 
\usepackage{xcolor}       
\usepackage{hyperref}	  
\hypersetup{colorlinks=false,linkcolor=black,citecolor=blue,breaklinks={true}} 
\usepackage[normalem]{ulem}

\usepackage{helvet} 
\usepackage{orcidlink} 
\usepackage{xurl}

\usepackage{graphicx}
\usepackage{sidecap}
\usepackage[font=footnotesize,labelfont=bf,justification=justified]{caption}
\graphicspath{{./FIG/}}
\DeclareGraphicsExtensions{.eps,.png,.jpg,.jpeg,.bmp,.gif,.pdf}

\usepackage{tabularx,booktabs}	
\usepackage{multirow}	
\usepackage{eurosym}

\usepackage{amsmath}      		
\usepackage{amssymb}	  		
\usepackage{bm}		      		
\usepackage{physics}	  		
\usepackage{mathtools}			
\usepackage{dsfont}		  		
\usepackage{soul}				
\usepackage{relsize}			

\usepackage{amsthm}

\theoremstyle{definition}

\usepackage[right]{lineno} 
\usepackage[resetlabels]{multibib}

\usepackage{authblk}                

\makeatletter
\renewcommand{\maketitle}{\bgroup\setlength{\parindent}{0pt}
\begin{flushleft}
  \textbf{\@title}
  
  \@author
\end{flushleft}\egroup}
\makeatother 

\usepackage{eurosym}

\title{\Large Rethinking the green power grid for stability, not just for climate} 

\author[1,2,*]{Yiming Wang} 
\author[1,2]{Arthur~N.~Montanari} 
\author[1,2,3,4]{Adilson E. Motter} 

\affil[1]{Center for Network Dynamics, Northwestern University, Evanston, IL 60208, USA} 
\affil[2]{Department of Physics and Astronomy, Northwestern University, Evanston, IL 60208, USA} 
\affil[3]{Northwestern Institute on Complex Systems, Northwestern University, Evanston, IL 60208, USA} 
\affil[4]{Department of Engineering Sciences and Applied Mathematics, Northwestern University, Evanston, IL 60208, USA} 
\affil[*]{Correspondence: yiming.wang2@northwestern.edu} 
\date{}

\begin{document}
\maketitle
\setcounter{footnote}{1}
\linenumbers


\bigskip\noindent
\textbf{Introduction}

\noindent
In March 2026, the European Network of Transmission System Operators for Electricity (ENTSO-E)  released its official report on the April 28, 2025, power outage in Spain and Portugal \cite{entsoe2025blackout}. What started as the disconnection of a few substations near Granada, Badajoz, and Seville quickly cascaded into a blackout across the whole Iberian Peninsula. Full restoration took almost 23  hours, a period during which more than 780 commercial flights were canceled, digital traffic dropped 83\%, and medical care disruptions resulted in at least eight deaths. In total, over 50 million people experienced a stark glimpse of modern society's dependence on electricity, which amounted to an estimated economic loss of \euro 1.6 billion. 
Significantly, at the time of the event, the Spanish grid was
operating with a 77.2\% renewable share 
\cite{entsoe2025blackout}, from solar (56.4\%), wind (10.2\%), and hydropower (10.6\%).
Major outages have occurred around the world under a wide range of generation mixes and have consistently prompted revisions of operational and security standards \cite{Pourbeik2006_Anatomy}. However, as renewable penetration increases, some of the underlying mechanisms governing power and voltage regulation as well as fault dynamics change, creating new challenges and opportunities for grid stability.

In an age marked by advanced sensing, communication, and control technologies, it might be surprising that we still experience wide-reaching power outages. Over the past decade, large-scale blackouts under high renewable penetration have occurred in the United States, 
Australia, 
and the United Kingdom. 
Diagnostic research on large-scale power system failures often points to aging grid infrastructure \cite{Pourbeik2006_Anatomy,yangyang}, the variability of intermittent renewable sources \cite{puertorico2025}, and reduced stability resulting from the displacement of conventional power plants \cite{IRENA2024_RePGC}. 
These diagnoses, taken together, reveal a deeper issue: 
Regulatory policymaking has been permissive of retrofitting and short-term operational fixes to accommodate renewable integration, rather than proactively mandating the adoption of renewable-based technologies designed to support grid stability.
Advances in generation technologies and power electronics now enable renewable resources to provide fast dynamic responses and actively enhance grid reliability, but these capabilities remain underutilized.

In this commentary, informed by the Iberian blackout, we argue that as wind and solar generation become dominant, renewable plants must be designed and operated not only to meet climate goals on decarbonization and reduced fossil fuel imports, but also as active assets in maintaining power system stability. 
This shift in perspective is crucial, as outages in modern power systems may no longer develop through sequential domino-like failures, but instead through sudden, system-wide disruptions akin to avalanches.

\bigskip\noindent
\textbf{From Dominoes to Avalanches}

\noindent
An unintended consequence of the rapidly changing energy landscape is that the time window once available to halt imminent blackouts is narrowing. In traditional power grids based on fossil-fuel, hydro, and nuclear plants, large-scale outages often begin as small localized failures that spread gradually over time. For example, an overloaded line could take tens of seconds to minutes to overheat before being eventually disconnected by protective devices, only then triggering a redistribution of power flows and potentially a cascade of secondary failures. 
This classic ``domino-like'' mechanism, exemplified by the 2003 US-Canada blackout \cite{Pourbeik2006_Anatomy}, is primarily governed by power-balance constraints and thermal accumulation, providing operators with
a brief but critical time window to intervene.
Today, in grids increasingly dominated by solar and wind generation, this window is becoming narrower. 
A contributing factor is that high-renewable integration can shift the dominant driving force of cascading failures toward faster dynamical processes. 
This shift is substantiated by evidence from the ENTSO-E investigation report on the Iberian blackout \cite{entsoe2025blackout}, which reveals that the disconnection of critical lines was not driven by the gradual thermal overloads typical of power-flow redistribution, but rather by near-instantaneous frequency excursions and out-of-step protection trips.
Renewable technologies based on power electronic inverters can adjust their output within seconds. While this responsiveness is valuable for counteracting price fluctuations in electricity markets, it also causes rapid power-flow imbalances and modifies the grid's inductive and capacitive behavior faster than regulation mechanisms can compensate for. The associated change in frequency response is ultimately determined by the displacement of the stabilizing inertia that would be provided by conventional generators. As a result, faster frequency responses can trigger protective actions on timescales far shorter than those associated with thermal overloads, leading to a qualitatively different cascade evolution. A recent analysis on the 2022 Puerto Rico blackout \cite{puertorico2025}, for example, demonstrated that high renewable integration can exacerbate the scale of catastrophic outages. 
Under such volatile conditions, even a localized disturbance, such as the tripping of a generator, can cause sharp fluctuations that propagate throughout the system in a timescale of seconds or less.

\begin{figure*}[t]
    \centering
    \includegraphics[width=\textwidth]{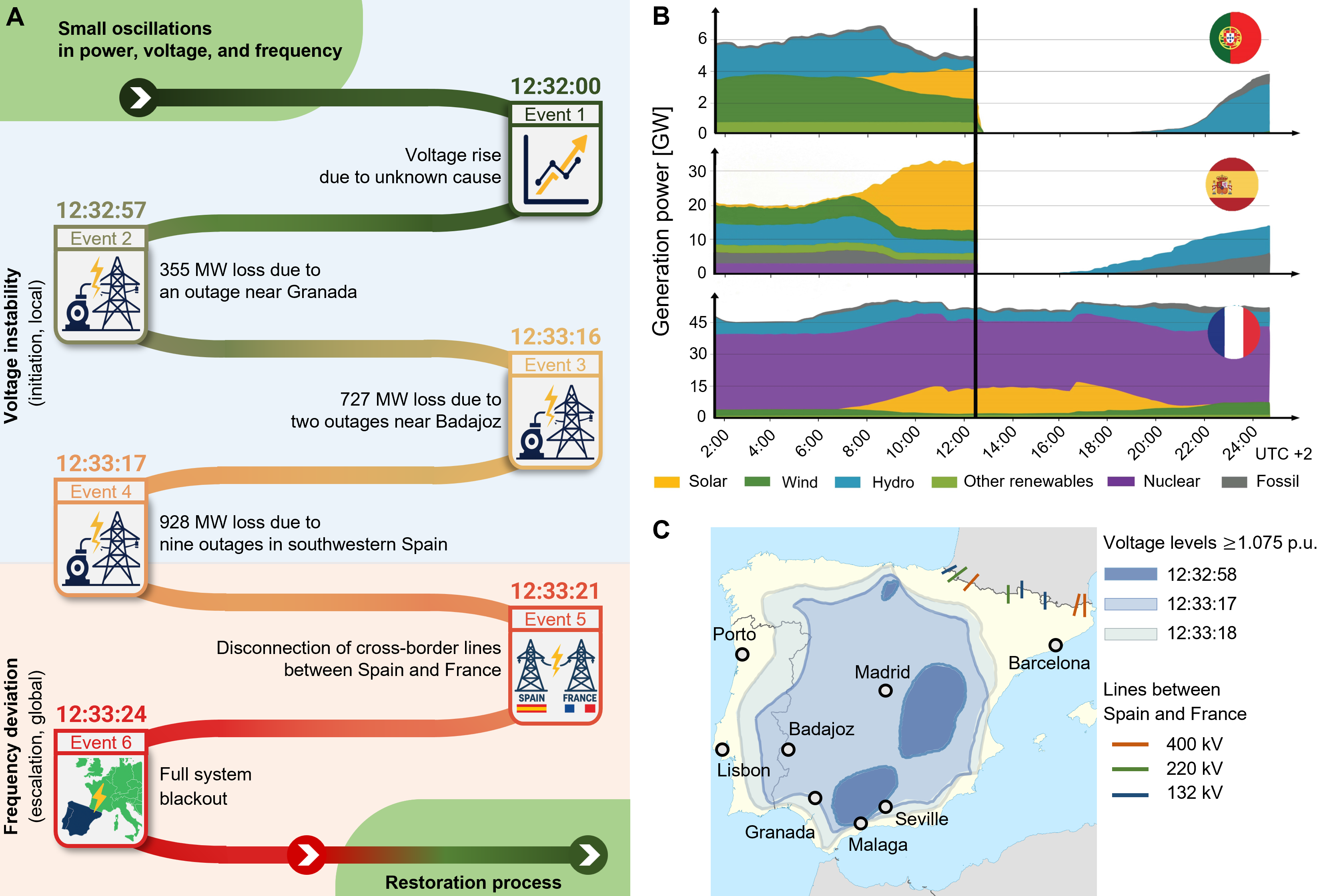}
    \caption{\label{fig.genmix} 
            \textbf{Overview of the 2025 Iberian blackout.}\\[0.2em]
            \textbf{(A)}~Timeline of major events leading up to the blackout. The initial generation outages were triggered by voltage instabilities, after which frequency disturbances rapidly propagated throughout the continental European system, resulting in the disconnection of the Spain-France cross-border lines.\\
            \textbf{(B)}~Generation mix in Portugal, Spain, and France (top to bottom) on the day of the blackout. The Iberian Peninsula has a high share of renewables compared with the predominance of nuclear in France. \\ 
            \textbf{(C)}~Voltage equipotential levels at different stages of the blackout, showing that abnormal voltage levels were largely confined to Spain and Portugal. 
            The information is based on public data from the transmission operators \cite{redeletricareport,spain,france}.}
\end{figure*}

In the morning before the 2025 Iberian blackout, oscillations in power, voltage, and frequency were observed across continental Europe but were mitigated by 12:30 p.m~(UTC+2) through existing control protocols\cite{entsoe2025blackout,redeletricareport}. However, at 12:32 p.m., voltage levels began to rise unexpectedly; 
ENTSO-E cites the loss of 500 megawatts of generation from solar and wind plants as a probable cause \cite{entsoe2025blackout}, though the evidence remains inconclusive due to insufficient high-resolution measurement data.
Less than a minute later, while voltage levels were still within the acceptable range, a solar and wind generation facility supplying 355 megawatts near Granada inadvertently tripped, likely due to a misconfigured transformer \cite{entsoe2025blackout,redeletricareport}. The outage pushed voltage levels even higher and was soon followed by the disconnection of at least ten other renewable generators, resulting in a total generation loss of 2{,}000 megawatts.
This perturbation, originating in southwestern Spain, reached the French border\textemdash more than 700 kilometers away\textemdash in less than four seconds. The resulting frequency deviations, and consequent loss of synchronization, triggered the disconnection of several cross-border lines, separating the two countries. The entire power grid of Spain and Portugal immediately collapsed, leading to a power loss of 38{,}000 megawatts (see Fig.~\ref{fig.genmix} for a timeline of events). Overall, there was little time for intervention as the regional events escalated. 

\bigskip\noindent
\textbf{Initiation versus Escalation}

\noindent
The assessment of ENTSO-E and \textit{Red Eléctrica de España}, the Spanish transmission system operator, is that a lack of inertia was not a root cause of the outage, attributing it instead to voltage instabilities linked to non-compliant reactive power control by conventional power plants \cite{entsoe2025blackout,redeletricareport}.
Although rising voltage levels are the identified cause for the initial generation losses in southwestern Spain (events 1-4 in Fig.~\ref{fig.genmix}A), the disconnection between Spain and France was ultimately caused by frequency disturbances (event 5).
Notably, measurement data indicate that voltage instabilities remained largely confined to the Iberian Peninsula~\cite{entsoe2025blackout}, whereas frequency disturbances spread across the whole continental European grid. For instance, deviations of about 0.1 Hz were recorded in eastern European cities like Riga (more than 2,500 kilometers away) shortly before the switches at the Spain–France border were activated \cite{radar2025}.
This dichotomy between voltage and frequency instabilities is well established in power systems theory.  
Voltage stability is governed primarily by local reactive power balances and therefore manifests as a geographically localized phenomenon, whereas frequency stability is tied to the synchronous operation of generators and the global balance between generation and load, enabling frequency disturbances to propagate rapidly across the whole system.
The Iberian blackout thus reflects two spatiotemporal layers: the initial failures driven by voltage instabilities, and the subsequent failures driven by frequency excursions.
Here, we focus mainly on the latter, which has received less attention and is the process through which 
the system’s frequency response determines whether a local disturbance triggers a wide-area collapse.

Frequency response is largely influenced by the system inertia. 
At the time of the blackout, the inertia of the Iberian system, expressed as kinetic energy per rated power, was estimated\cite{entsoe2025blackout} to be 2.19--2.69~s. This value is close to the lower bound of the projected 2025 range of 2.5--4~s for the continental European system\cite{ENTSOE_inertia}, 
reflecting the limited inertia support and system services required by current market and regulatory arrangements in Europe~\cite{EU_Commision}. The fact that prevailing policies do not yet mandate the provision of adequate inertia in high-renewable conditions can amplify the system
 vulnerability to frequency fluctuations, especially when combined with lagged transmission upgrades.
Given the key role of renewables toward achieving zero carbon emissions, it is essential to invest in the comprehensive redesign of the transmission grid and control protocols, which currently rely too heavily on regional planning. These investments have the potential to address other barriers to the green-energy transition, including easing the backlog in the interconnection queues, where proposed energy projects wait for approval to connect to the grid \cite{armstrong2024can}.

\bigskip\noindent
\textbf{Making Form Fit Function}

\begin{figure*}[t]
    \centering
    \includegraphics[width=0.98\textwidth]{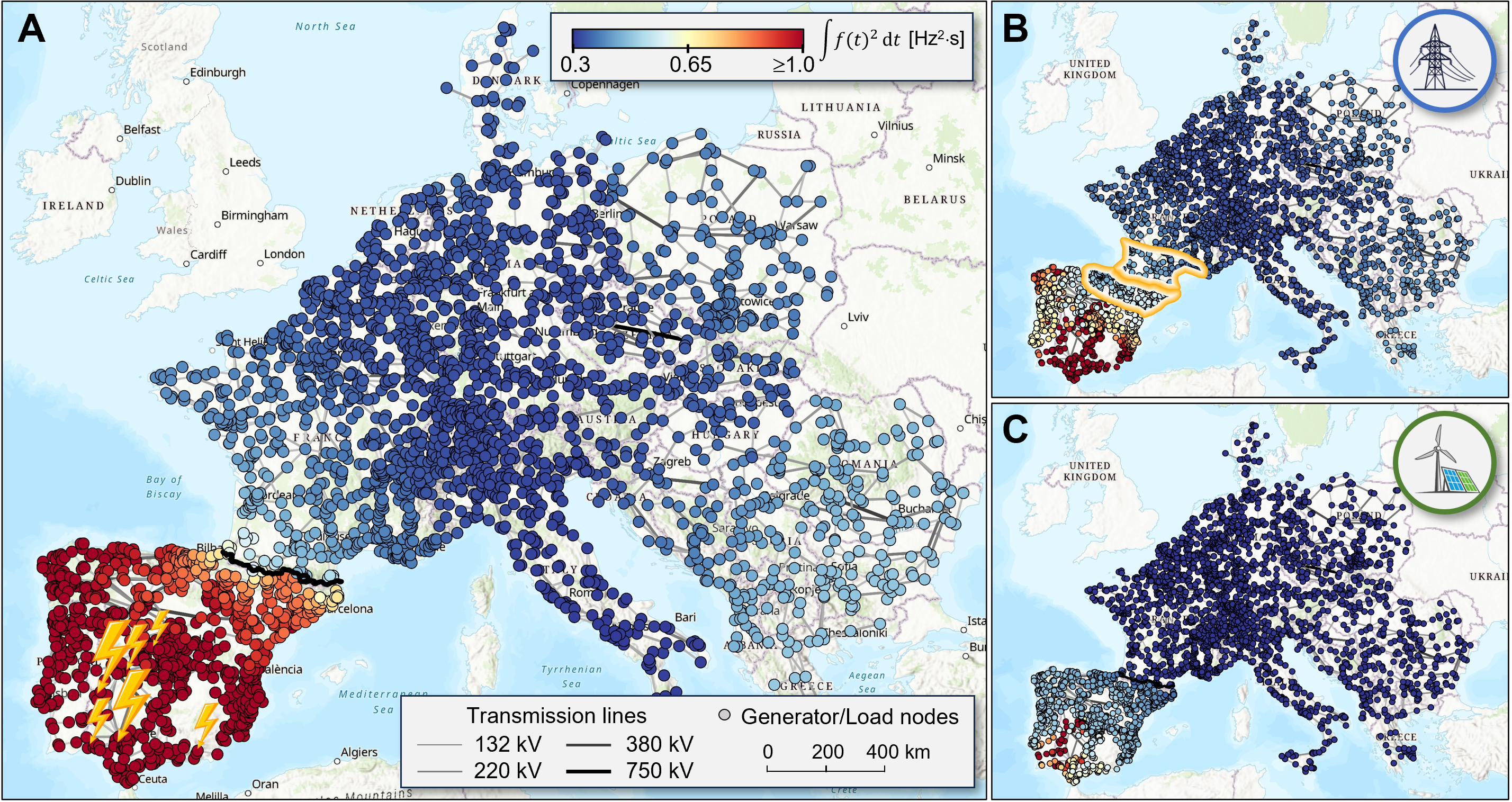}
    \captionsetup{justification=justified}
  \caption{\textbf{Cumulative frequency deviation building up to the 2025 Iberian blackout.}\\[0.2em]
  Simulated frequency dynamics using a continental-scale European power grid model \cite{SimulatoinModel} and the generation mix at the time of the blackout.
  The lightning strikes mark the initial faults, corresponding to a 1,916 megawatts generation loss in southwestern Spain. The model implementation and simulation details are provided in the accompanying code. \\
  \textbf{(A)}~Simulation using nominal parameters. A pronounced frequency cliff emerges at the Spain-France border (black line), with frequency fluctuations surpassing 1 Hz within the peninsula. This behavior reproduces the loss of synchronism between Spain and France reported by ENTSO-E \cite{entsoe2025blackout}. \\
  \textbf{(B)}~Simulation under transmissions upgrades. Frequency fluctuations are strongly suppressed by mitigation strategies based on reinforcing cross-border connectivity, shown for the addition of ten new transmission lines between the highlighted regions. \\
  \textbf{(C)}~Simulations under synthetic inertia upgrades. Fluctuations are also strongly suppressed when the inertia contribution from solar and wind generators relative to the whole-system inertia is increased to 59.4\% across the Iberian Peninsula (up from 4.6\% in panel A). In this scenario, solar and wind units are modeled as grid-forming inverters with virtual synchronous machine control, providing synthetic inertia to the system\cite{rosso2021gfm}. 
  The synthetic inertia of inverter-based generators is uniformly increased, while remaining within nominal parameter range specified in the ENTSO-E technical report on grid-forming capability \cite{entsoe2024gfc}. Frequency deviations across buses show a robust, monotonic decrease with the inertia scaling factor.
}
\label{fig.simulation}
\end{figure*}

\noindent
Because in the European grid all power generators must operate synchronously near 50 Hz under normal conditions, deviations from this frequency can be used to quantify the impact of a disturbance to the grid. Simulations show that, following the generation loss in southwestern Spain, a striking \textit{frequency cliff} appears at the Spain–France border (Fig.~\ref{fig.simulation}A). On the Spanish side, frequency deviations swell into a nationwide blackout, with magnitudes well beyond 1 Hz; on the French side, the disturbances propagated rapidly across the European system but remained largely subdued. Thus while the perturbation spread far into the entire Iberian Peninsula, this cliff held firm and did not ripple deeper into continental Europe (although weak long-range deviations can be observed in Southeast Europe). 
This effect is counterintuitive since, in large synchronous power systems, frequency disturbances are generally expected to attenuate gradually with electrical distance from the initiating event rather than exhibiting an abrupt change; indeed, the observed behavior occurred while the Spanish and French systems were still electrically connected, before the cross-border lines were disconnected.
An intuitive approach to prevent wide-area impact is to strengthen the connectivity across the Spain-France border, and this is confirmed by the simulated addition of ten new 380 kV lines within a 250-kilometer radius of the border (Fig.~\ref{fig.simulation}B). 
Importantly, a submarine power cable is being built between Spain and France, which is expected to increase the cross-boundary capacity from 2{,}800 to 5{,}000 megawatts by 2028 at the cost of \euro 1.6 billion \cite{submarine}.

A key question is the extent to which renewables themselves can be designed and operated to boost grid resilience.  
Traditional power grids, relying on fossil, hydro, and nuclear power plants, are built with inherently strong frequency control, since their large, heavy turbines provide rotational inertia that resists sudden changes to the grid's operating frequency. Solar and wind generators, on the other hand, do not naturally provide inertia. Increasing the share of solar and wind generation can thus inadvertently lead to increased frequency fluctuations. Yet, as electromechanical systems, these renewable generators can be engineered to synthetically emulate inertia through electrical means, including energy storage-and-release battery systems, capacitor banks, and grid-forming inverters. Therefore, adjusting the inertia of renewables in the Iberian Peninsula is another promising direction to mitigate the amplification of disturbances (Fig.~\ref{fig.simulation}C). The effectiveness of this approach has been demonstrated for other systems, such as in a pilot implementation within the Australian power system \cite{ESCRI}.

The optimal inertia of each generator in the system generally depends on the grid's operational state. In contrast with the static inertia provided by conventional generators, the synthetic inertia in solar and wind generators can be actively controlled with the support of advanced power electronics and control systems. This highlights a key advantage of renewable-rich power grids: the ability to dynamically adjust the level of inertia based on real-time system needs, thereby improving grid resilience beyond what is possible with conventional generation alone. 
Moreover, this inertia can be optimized, and sometimes even reduced over short-time periods, in order to absorb frequency oscillations and quickly resynchronize in the event of disturbances \cite{virtualinertia}. 

\begin{figure}[t]
    \centering
    \includegraphics[width=0.8\columnwidth]{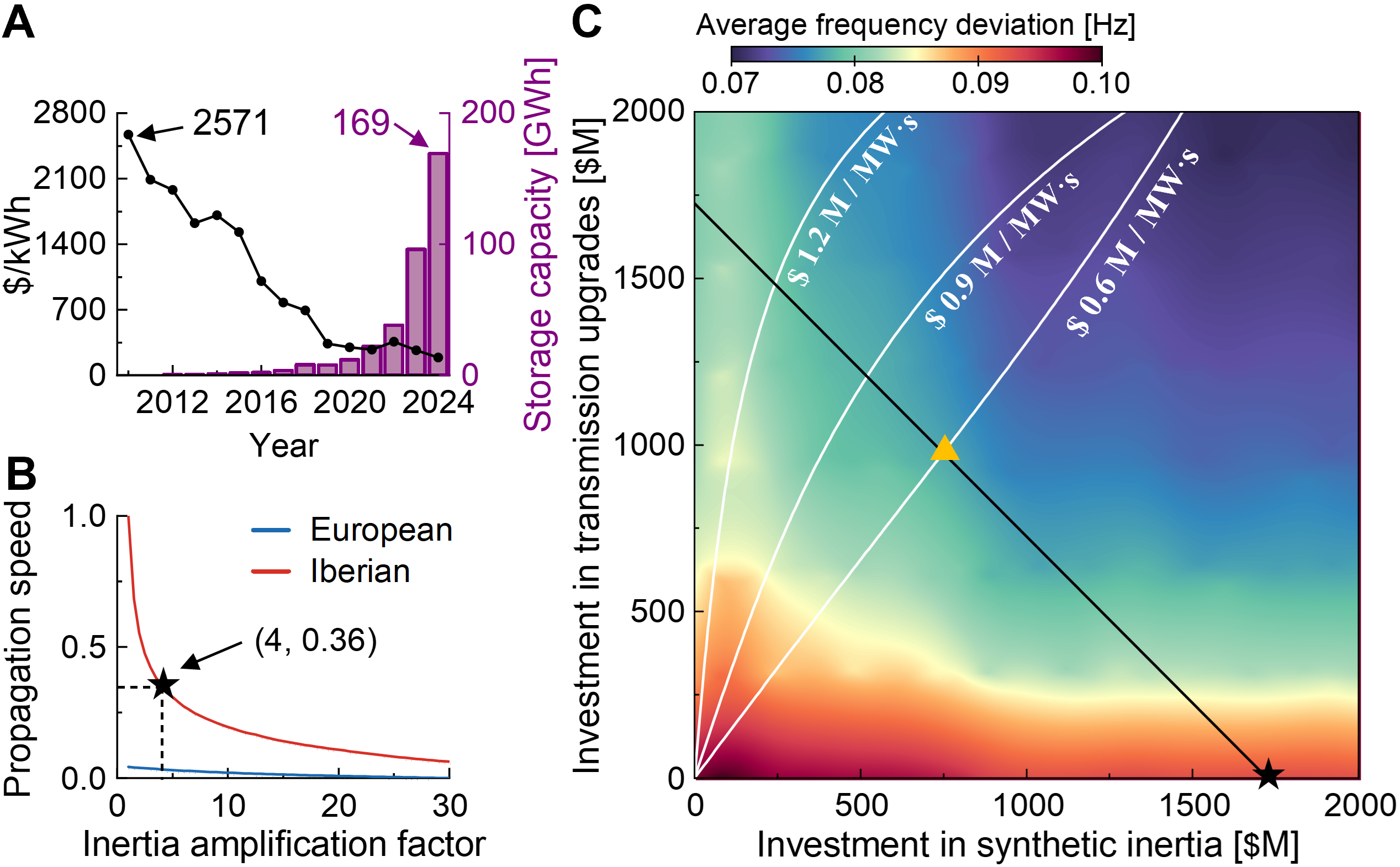}
    \captionsetup{justification=justified}
    \caption{
     \textbf{Cost trade-offs between synthetic inertia and transmission network upgrades.}\\[0.2em]
    \textbf{(A)}~Historical cost decrease and deployed capacity increase of utility-scale battery energy storage systems, based on IRENA data~\cite{IRENA2024_RePGC}. \\
    \textbf{(B)}~Propagation speed of frequency disturbances in the Iberian system as a function of the amplification of inertia, both relative to the current baseline. \\
    \textbf{(C)}~Average frequency deviation as a function of battery deployment costs for synthetic inertia and cross-border transmission upgrade costs, where the simulated average is taken over the first 2 seconds from the triggering event of the Iberian blackout. The heatmap corresponds to estimated synthetic inertia costs of 0.6 million dollars/MW·s and overhead transmission lines costs of 1 million dollars/km  %
    (using the same upgrade rules as in Fig.~\ref{fig.simulation}, with distances between buses measured along straight lines).
    The black star marks the total cost of increasing inertia by a factor of four (as highlighted in panel B), while the black line indicates the corresponding locus of equal investment cost and the orange triangle marks the corresponding smallest frequency deviation point.
    The white lines show the optimal investment path to minimize frequency deviations considering different prices per unit of inertia, ranging from 0.6 to 1.2 million dollars/MW·s.
    Following the original sources, all prices are reported in dollars.
}
    \label{fig:cost}
\end{figure}

As the costs of battery systems continue to decrease rapidly \cite{IRENA2024_RePGC}, the deployment of synthetic inertia alongside renewable technologies has become increasingly accessible in recent years (Fig.~\ref{fig:cost}A).
Increasing inertia by a factor of four within the Iberian Peninsula cuts propagation speed to 36\% of its original value, aiding fault localization and extending operator response time (Fig.~\ref{fig:cost}B). 
However, as shown in Fig.~\ref{fig:cost}C, the most cost-effective path to strengthening grid resilience is the coordinated investment in both synthetic inertia and transmission upgrade. For example, while investing 1.7 billion dollars exclusively in synthetic inertia can reduce frequency deviations by 9\% (black star), the same financial investment can lead to an improvement of 23\% (orange triangle) when combining inertia deployment and transmission upgrade.

\bigskip\noindent
\textbf{Turning Change into  Opportunity}

\noindent
The Iberian blackout, much like the February 2021 outages in Texas, has once again placed renewable energy under intense scrutiny. While it is well understood that the intermittency of solar and wind generation is not the root of failures \cite{renewable_vulnerability}, the role of the renewable transition in the broader context of energy system resilience warrants careful consideration. 
Developing mitigation plans for cascades requires an understanding of the many interacting factors that underlie the propagation of such failures, including heterogeneous generation sources, transmission constraints, and variable load demand.
To this end, the Iberian event can serve as a good case study to isolate effects that would be conflated in other systems.  At the time of the blackout, the peninsula had 70\% of its generation from solar and wind, while France had only about 25\% (Fig.~\ref{fig.genmix}B). 
Thus, not only is France more strongly interconnected with the rest of the continental European grid, but it also has a substantially higher inertia than Spain and Portugal (six times larger for the conditions in Fig.~\ref{fig.simulation}A) due to its large share of nuclear energy.
As shown in Fig.~\ref{fig.simulation}B, strengthening the connection between Spain and France would enable the Iberian system to ``import'' grid stability from France rather than ``export'' instability. 
While the largest frequency deviations were mainly limited to the Iberian system, strengthening interconnections can further mitigate the propagation of disturbances into the rest of the European grid, yielding mutual stability benefits for both Spain and France. Such strongly heterogeneous spatial distribution of renewable generation is not unique to Europe. For example, in the U.S., the state of Iowa currently generates around 65\% of its power from solar and wind, while the neighboring state of Missouri generates only 10\% \cite{EIA2025Electricity}.

In complex systems like modern power grids, the occurrence of localized failures---whether due to extreme weather, equipment breakdown, or  operational errors---is often inevitable. What matters most is whether one can prevent these failures from cascading into large-scale outages. 
For example, the observed asymmetry in frequency responses relative to the Spain-France border highlights the need of accelerating the transition toward active grid support from renewables.
The properly planned integration of renewables can play a central role in doing just that.
Advanced technologies such as grid-forming inverters (providing synthetic inertia), phasor measurement units (enabling precise, real-time sensing and system observability), and synchronous condensers (facilitating dynamic voltage control, damping, and reactive power compensation) are powerful tools to contain failures before they escalate. 
In Texas, such innovations are already being considered for the containment of frequency fluctuations \cite{huang2024ercot}. In Spain, the Ministry for the Ecological Transition and the Demographic Challenge has also proposed measures to upgrade renewable systems with additional energy storage \cite{crisis_electricidad_2025}.
However, as recent blackouts suggest, cutting-edge renewable technologies will only reach their full potential in both enhancing system resilience and fulfilling their core promise of clean energy generation if their integration is accompanied by a system-wide coordinated upgrade of the transmission infrastructure.

\smallskip\noindent
\textbf{Data and materials availability.}
Code, data, and supporting materials are available at \url{https://github.com/YimingSci/EUPG-Renewables}.

\smallskip\noindent
\textbf{Declaration of interests.}
The authors declare no competing interests.

\smallskip\noindent
\textbf{Acknowledgments.}
This work benefitted from resources from NSF grant No. DMS-2308341.

\smallskip\noindent
\textbf{Author contributions statement.}
Y.W., A.N.M., and A.E.M. conceptualized the research;
Y.W. curated the data, performed the numerical simulations, and analyzed the data;
all authors contributed to the writing of the manuscript, interpretation of the results, and editing of the final version of the paper.

\begin{footnotesize}

\end{footnotesize}

\clearpage\newpage


\begin{thebibliography}{10}
\setlength{\itemsep}{0pt}  %
\setlength{\parsep}{0pt}   %



















\bibitem{entsoe2025blackout}
European Network of Transmission System Operators for Electricity (ENTSO-E) (2026). Iberian blackout on 28 April 2025.
\url{https://eepublicdownloads.blob.core.windows.net/public-cdn-container/clean-documents/Publications/2025/iberian-blackout/Final%20Report%20on%20the%20Grid%20Incident%20in%20Spain%20and%20Portugal%20on%2028%20April%202025.pdf}.

\bibitem{Pourbeik2006_Anatomy} 
Pourbeik, P., Kundur, P.S., and Taylor, C.W. (2006). The anatomy of a power grid blackout: root causes and dynamics of recent major blackouts. IEEE Power Energy Mag. 4, 22--29.
\url{https://doi.org/10.1109/MPAE.2006.1687814}.

\bibitem{yangyang}
Yang, Y., Nishikawa, T., and Motter, A.E. (2017). Small vulnerable sets determine large network cascades in power grids. Science 358, eaao3184.
\url{https://doi.org/10.1126/science.aan3184}.

\bibitem{puertorico2025}
Xu, L., Lin, N., Poor, H.V., Xi, D., and Perera, A.D. (2025). Quantifying cascading power outages during climate extremes considering renewable energy integration. Nat. Commun. 16, 2582.
\url{https://doi.org/10.1038/s41467-025-57565-4}.


\bibitem{IRENA2024_RePGC}
International Renewable Energy Agency (2025). Renewable power generation costs in 2024. 
\url{https://www.irena.org/-/media/Files/IRENA/Agency/Publication/2025/Jul/IRENA_TEC_RPGC_in_2024_2025.pdf}.

\bibitem{redeletricareport}
Red Eléctrica de España (REE) (2025). Blackout in Spanish Peninsular electrical system the 28th of April 2025.
\url{https://d1n1o4zeyfu21r.cloudfront.net/WEB_Incident_%2028A_SpanishPeninsularElectricalSystem_18june25.pdf}.

\bibitem{spain}
REE. National electricity demand tracking in real time.
\url{https://demanda.ree.es/visiona/peninsula/nacionalau/total}.

\bibitem{france}
RTE France. éCO$_{2}$mix -- power generation by energy source.
\url{https://www.rte-france.com/en/eco2mix/power-generation-energy-source}.

\bibitem{radar2025}
Gridradar (2025). Blackout -- Iberian Peninsula.
\url{https://gridradar.net/en/blog/post/blackout-iberian-peninsula}.


\bibitem{ENTSOE_inertia}
ENTSO-E (2020). System dynamic and operational challenges. 
\url{https://eepublicdownloads.blob.core.windows.net/public-cdn-container/tyndp-documents/IoSN2020/200810_IoSN2020_Systemdynamicandoperationalchallenges.pdf}.

\bibitem{EU_Commision}
European Commission (2025). Assessment of policy options for securing inertia.
\url{https://www.artelys.com/app/uploads/2025/09/assessment-of-policy-options-for-securing-inertia.pdf}.

\bibitem{armstrong2024can}
Armstrong, L., Canaan, A., Knittel, C., Metcalf, G., and Schittekatte, T. (2024). Can federal grid reforms solve the interconnection problem?. Science 385, 31--33.
\url{https://doi.org/10.1126/science.ado9254}.


\bibitem{SimulatoinModel}
Tyloo, M., Pagnier, L., and Jacquod, P. (2019). The key player problem in complex oscillator networks and electric power grids: resistance centralities identify local vulnerabilities. Sci. Adv. 5, eaaw8359.
\url{https://doi.org/10.1126/sciadv.aaw8359}.

\bibitem{rosso2021gfm}
Rosso, R., Wang, X., Liserre, M., Lu, X., and Engelken, S. (2021). Grid-forming converters: control approaches, grid-synchronization, and future trends -- a review. IEEE Open J. Ind. Appl. 2, 93–109.
\url{https://doi.org/10.1109/OJIA.2021.3074028}.

\bibitem{entsoe2024gfc}
ENTSO-E (2024). Grid forming capability of power park modules: first interim report on technical requirements.
\url{https://eepublicdownloads.entsoe.eu/clean-documents/Publications/SOC/20240503_First_interim_report_in_technical_requirements.pdf}.

\bibitem{submarine}
REE (2025). The electrical connection between France and Spain in the Bay of Biscay has achieved a new milestone with the laying of the first stone for the converter station at Cubnezais.
\url{https://www.ree.es/sites/default/files/2025-05/PR_The%20electrical%20connection%20between%20France%20and%20Spain%20in%20the%20Bay%20of%20Biscay%20has%20achieved%20a%20new%20milestone.pdf}.

\bibitem{ESCRI}
ElectraNet and Australian Renewable Energy Agency (2021). ESCRI-SA battery energy storage project operational report \#4. 
\url{https://arena.gov.au/assets/2021/04/escri-sa-battery-energy-storage-report-4.pdf}.

\bibitem{virtualinertia}
Fritzsch, J., and Jacquod, P. (2024). Stabilizing large-scale electric power grids with adaptive inertia. PRX Energy 3, 033003.
\url{https://doi.org/10.1103/PRXEnergy.3.033003}.

\bibitem{renewable_vulnerability}
Zhao, J., Li, F., and Zhang, Q. (2024). Impacts of renewable energy resources on the weather vulnerability of power systems. Nat. Energy 9, 1407--1414.
\url{https://doi.org/10.1038/s41560-024-01652-1}.

\bibitem{EIA2025Electricity}
U.S. Energy Information Administration. Electricity data browser -- state electricity profiles.
\url{https://www.eia.gov/electricity/data.php}.

\bibitem{huang2024ercot}
Electric Reliability Council of Texas (2024). ERCOT advanced grid support inverter-based energy storage system assessment and adoption discussion.
\url{https://www.ercot.com/files/docs/2024/07/09/2024_07_ERCOT_IBRWG_ERCOT%20Advanced%20Grid%20Support%20Inverter-based%20ESRs%20Assessment%20and%20Adoption%20Discussion_v1_.pdf}.

\bibitem{crisis_electricidad_2025}
Comité para el Análisis de las Circunstancias que Concurrieron en la Crisis de Electricidad del 28 de Abril de 2025. (2025). Versión no confidencial del informe del comité para el análisis de las circunstancias que concurrieron en la crisis de electricidad del 28 de abril de 2025.
\url{https://www.lamoncloa.gob.es/consejodeministros/resumenes/Documents/2025/Informe-no-confidencial-Comite-de-analisis-28A.pdf}.











\end{thebibliography}
\end{document}